\documentclass[11pt,journal,draftclsnofoot,a4paper,oneside,onecolumn]{IEEEtran}

\usepackage{cite}
\usepackage{graphicx}
\usepackage{psfrag}
\usepackage{subfigure}
\usepackage{url}
\usepackage{stfloats}
\usepackage{amsmath}
\usepackage{amssymb}
\usepackage{array}

\newcounter{step}
\newlength{\totlinewidth}
\newenvironment{algorithm}{%
  \rule{\linewidth}{1pt}
  \begin{list}{}%
    {\usecounter{step}%
      \settowidth{\labelwidth}{\textbf{Step 2:}}%
      \setlength{\leftmargin}{\labelwidth}%
      \setlength{\topsep}{-2pt}%
      \addtolength{\leftmargin}{\labelsep}%
      \addtolength{\leftmargin}{2mm}%
      \setlength{\rightmargin}{2mm}%
      \setlength{\totlinewidth}{\linewidth}%
      \addtolength{\totlinewidth}{\leftmargin}%
      \addtolength{\totlinewidth}{\rightmargin}%
      \setlength{\parsep}{0mm}%
      \raggedright}}%
  {\end{list}%
  \rule{\linewidth}{1pt}}
\newcounter{substep}

\newenvironment{subalgorithm}{%
  \begin{list}{}%
    {\usecounter{substep}%
      \setlength{\topsep}{0pt}%
      \setlength{\parskip}{0pt}%
      \setlength{\partopsep}{0pt}%
      \settowidth{\labelwidth}{\textbf{Step iii:}}%
      \setlength{\leftmargin}{\labelwidth}%
      \addtolength{\leftmargin}{\labelsep}%
      \addtolength{\leftmargin}{-10mm}%
      \setlength{\rightmargin}{2mm}%
      \setlength{\totlinewidth}{\linewidth}%
      \addtolength{\totlinewidth}{\leftmargin}%
      \addtolength{\totlinewidth}{\rightmargin}%
      \setlength{\parsep}{0mm}%
      \raggedright}}%
  {\end{list}}

\newcommand{\indentcolumn}[1]{%
  \begin{list}{}{%
      \setlength{\topsep}{0pt}%
      \setlength{\parskip}{0pt}%
      \setlength{\partopsep}{0pt}%
      \setlength{\itemsep}{0pt}%
      \setlength{\labelwidth}{0pt}%
      \addtolength{\leftmargin}{0.3em}}%
  \item #1
    \vspace*{-1pt}
  \end{list}}

\newlength{\aligntop}
\newlength{\alignbot}
 \makeatletter
\renewenvironment{align}{%
  \vspace{\aligntop}
  \start@align\@ne\st@rredfalse\m@ne
}{%
  \math@cr \black@\totwidth@
  \egroup
  \ifingather@
    \restorealignstate@
    \egroup
    \nonumber
    \ifnum0=`{\fi\iffalse}\fi
  \else
    $$%
  \fi
  \ignorespacesafterend%
  \vspace{\alignbot}\par\noindent
} \makeatother

\begin{document}
%
%
\title{Approximate ML Decision Feedback Block Equalizer for Doubly Selective Fading Channels}
\author{Lingyang~Song, Rodrigo~C.~de~Lamare, Are~Hj{\o}rungnes, and Alister~G.~Burr
\thanks{Lingyang~Song is with School of Electrical Engineering and Computer Science, Peking University, Beijing, China (e-mail: \protect\url{lingyang.song@pku.edu.cn}).}
\thanks{Are~Hj{\o}ungnes is with UniK-University Graduate Center, University of Oslo, Norway
(e-mail: \protect\url{ arehj@unik.no}).}
\thanks{Rodrigo~C.~de~Lamare and Alister~G.~Burr are with the Department of Electronics,
University of York, York, UK (e-mail:
\protect\url{rcdl500@ohm.york.ac.uk;}
\protect\url{alister@ohm.york.ac.uk}).}
}
%
%
\maketitle 
\begin{abstract}
In order to effetively suppress intersymbol interference~(ISI) at
low complexity, we propose in this paper an approximate maximum
likelihood~(ML) decision feedback block equalizer~(A-ML-DFBE) for
doubly selective (frequency-selective, time-selective) fading
channels. The proposed equalizer design makes efficient use of the
special time-domain representation of the multipath channels through
a matched filter, a sliding window, a Gaussian approximation, and a
decision feedback. The A-ML-DFBE has the following features: 1) It
achieves performance close to maximum likelihood sequence
estimation~(MLSE), and significantly outperforms the minimum mean
square error~(MMSE) based detectors; 2) It has substantially lower
complexity than the conventional equalizers; 3) It easily realizes
the complexity and performance tradeoff by adjusting the length of
the sliding window; 4) It has a simple and fixed-length feedback
filter. The symbol error rate~(SER) is derived to characterize the
behaviour of the A-ML-DFBE, and it can also be used to find the key
parameters of the proposed equalizer. In addition, we further prove
that the A-ML-DFBE obtains full multipath diversity.
\end{abstract}
\section{INTRODUCTION}
\PARstart{W}{ireless} communications often suffer from severe
inter-symbol interference (ISI) due to doubly selective fading. In
order to suppress the channel distortion, channel equalization
techniques are essential, and indeed have received considerable
attention for many years. Maximum a priori probability~(MAP)
equalization is the optimum equalization procedure in terms of
minimum symbol error rate (SER)~\cite{Proakis-Digital-Comms}, but
requires a prohibitive computational complexity for many
applications, being exponential in the channel length and
constellation size. Maximum likelihood sequence estimation (MLSE)
can obtain SER performance very close to MAP, but its complexity is
still extremely high~\cite{IEEE-Forney-MLSE}. As a result, many
sub-optimal, low-complexity equalization techniques have been
proposed, such as the popular minimum mean square error
decision-feedback equalizer, which is very effective in certain
multipath environments and has a complexity that is only dependent
on forward and backward filter lengths~\cite{IEEE-MMSE-DFE}.
However, there is a non-negligible performance loss of MMSE based
equalizers in comparison to
MLSE~\cite{Simon-Adaptive-Filter}~\cite{IEEE-Geert-Block-Equalization}.

Further still, while lots of research have been conducted on the
time-domain equalization, few works take the special form of the
channel representation into good account. Two properties of the
channel matrix in time domain are effectively utilized in this
paper: 1) The Toeplitz-like channel matrix significantly contributes
to the equalizer design; 2) The large number of zero elements
reduces the computational complexity. As a result, we propose a
robust \emph{approximate ML based decision feedback block equalizer}
(A-ML-DFBE) to combat ISI over doubly selective fading channels with
low computational complexity. The proposed equalizer exploits
substantial benefit from the special time domain representation of
the multipath channels by using a \emph{matched filter}, a
\emph{sliding window}, a \emph{Gaussian approximation}, and a
\emph{decision feedback}. The main ideas are firstly to subtract the
effect of the already-detected signals obtained from past decisions.
This can be treated as a decision feedback process. Secondly we
apply Gaussian
approximation~\cite{IEEE-Lingyang-SIC}--\hspace{-0.03em}\cite{IEEE-Tian-PDA}
to realize near maximum likelihood detection. The accuracy of this
procedure can be improved by adjusting the length of the sliding
window due to the central limit theorem. Consequently, a complexity
and performance trade-off can be realized, and a convergence in SER
performance can also be obtained by adjusting the length of the
sliding window.

Note that \cite{IEEE-Lingyang-SIC} and \cite{IEEE-Yugang-SGA} can be
used only for frequency flat fading channels, and
\cite{IEEE-Luo-PDA} aims to recover signals for multiuser systems.
Although in \cite{IEEE-Tian-PDA} a probabilistic data
association~(PDA) based equalizer is reported, there are several
major differences compared to the proposed approach: In
\cite{IEEE-Tian-PDA}, it requires to update the mean and the
variance for all detected symbols; many iterations have to be used
in order to make the performance converge; there is no feedback
process; and no matched filter is employed. In \cite{IEEE-BAD},
bidirectional arbitrated decision-feedback equalization~(BAD)
algorithm was presented which has complexity at least two times of
the MMSE-DFE but can achieve better performance. In
\cite{IEEE-Wornell-BDFE}, a class of block DFE is presented for
frequency domain equalization, but it assumes that the length of the
channel, forward filter, and backward filter are infinitely long
which is not practical. Besides, it requires large number of
iterations to make the performance converge, which increases the
system delay and the computational complexity.


The rest of the paper is organized as follows: In Section II, we
present the channel and signal models. The proposed A-ML-DFBE scheme
and complexity comparisons are discussed in Section III. The
performance is analyzed in Section IV. Simulation results are
presented in Section V. In Section VI, we draw the main conclusions.
The proof is given in the appendix.

\emph{\textbf{Notation}}: Boldface upper-case letters denote
matrices, boldface lower-case letters denote vectors,
$\mathbb{C}^{i\times{j}}$ and $\mathbb{R}^{i\times{j}}$ denote the
set of $i\times{j}$ complex and real matrices, respectively,
$(\cdot)^{T}$ stands for transpose, $(\cdot)^{*}$ denotes complex
conjugate, $(\cdot)^{H}$ represents conjugate transpose,
$\textbf{I}_i$ stands for an $i\times{i}$ identity matrix,
$\boldsymbol{\mathbb{E}}$ is used for expectation, $\texttt{var}$ is
used for variance, and $\|\textbf{x}\|^2=\textbf{x}^H\textbf{x}$.
\section{Channel and Signal Models}
\label{sec:PRELIMINARIES}
The doubly selective fading channel can be modeled using a finite
impulse response (FIR) filter
\begin{equation}
    H(z,t)=\sum_{k=0}^{L-1}h_k(t)\emph{z}^{-k},
  \label{Eq:SysModel}
\end{equation}
where $H(z,t)$ denotes the $z$ transformation at time $t$, $h_i(t)$
represents the $\emph{i}$-th path's channel coefficient, and the
length of the FIR filter is $L$. For simplicity, we only consider a
single input and single output system. The received signals can be
written in vector form as (for convenience, we drop the time index
for each transmission frame)
\begin{equation}
    \textbf{r}=\textbf{H}\textbf{s}+\textbf{n},
        \label{Eq:RecModel}
\end{equation}
where the received signals
$\textbf{r}=[\emph{r}_1,{\ldots},\emph{r}_{\emph{N}+\emph{L}-1}]^\emph{T}$,
$N$ is the length of $\textbf{s}$, transmitted signals
$\textbf{s}=[\emph{s}_1,{\ldots},\emph{s}_N]^\emph{T}$, and
$\textbf{n}=[\emph{n}_1,{\ldots},\emph{n}_{\emph{N}+\emph{L}-1}]^\emph{T}$
whose elements are independent samples of a zero-mean complex
Gaussian random variable with variance
$\sigma^{2}=\boldsymbol{\mathbb{E}}[|s_k|^2]/\texttt{SNR}$, in which
$\boldsymbol{\mathbb{E}}[|s_k|^2]$ represents the average power of
the transmitted symbols from constellation $\mathcal{A}$. In this
paper, we set $\boldsymbol{\mathbb{E}}[|s_k|^2]=1$. The time domain
representation of the doubly selective fading channel
$\textbf{H}\in\mathbb{C}^{(\emph{N}+\emph{L}-1)\times\emph{N}}$, can
be written as
$$
        \textbf{H}= \left[\begin{array}{ccccc}
    \emph{h}_1(0)&0&0&\ldots&0\\
    \emph{h}_2(0)&\emph{h}_1(1)&0&\ldots&\vdots\\
    \vdots&\emph{h}_2(1)&\emph{h}_1(2)&\ldots&0\\
    \emph{h}_\emph{L}(0)&\vdots&\emph{h}_2(2)&\ldots&\emph{h}_1(N-1)\\
    0&\emph{h}_\emph{L}(1)&\vdots&\ldots&\emph{h}_2(N-1)\\
    \vdots&\vdots&\vdots&\ldots&\vdots\\
    0&0&0&\ldots&\emph{h}_\emph{L}(N-1)\\
    \end{array}\right].
  \label{Eq:SliceJ}
$$
Note that $\textbf{H}$ has a structure similar to Toeplitz form, and
some form of guard interval is necessary to avoid inter-block
interference between the received signals
\cite{IEEE-Geert-Block-Equalization}. The symbols in
(\ref{Eq:RecModel}) can be recovered by
MLSE~\cite{Proakis-Digital-Comms}. Alternatively, they can also be
decoded in complex form using standard zero forcing (ZF) or MMSE
approaches, linear or decision feedback
equalization~\cite{IEEE-MMSE-DFE}.
\section{Description of the Proposed Method}
\label{sec:PRELIMINARIES}
\subsection{Approximate Maximum Likelihood Decision Feedback Block Equalizer (A-ML-DFBE)}
The proposed equalization algorithm can be summarized into three
steps: 1) Forward process, which builds up the forward filter by a
temporal sub matched filter; 2) Decision feedback process, which
cancels the interference by a fixed length backward filter, and 3)
Approximate ML process, which realizes the final signal detection by
the aid of Gaussian approximation. The detailed description of each
step is given below.
\subsubsection{Forward Process}
Supposing we start decoding $s_k$, a temporal sub matched
filter~(forward filter) is applied to~(\ref{Eq:RecModel})
\begin{equation}
    \textbf{H}_{k}^H\textbf{r}=\textbf{H}_{k}^H
    \textbf{H}\textbf{s}+\textbf{H}_{k}^H\textbf{n},
  \label{Eq:Hrec}
\end{equation}
where $\textbf{H}_{k}$ denotes the matrix of size
${N}\times\emph{L}_\emph{f}$, which is made of the entries in
$\textbf{H}$, from the $\emph{k}$-th column to the
$(\emph{k}+\emph{L}_\emph{f}-1)$-th column and from the $1$-st row
to the $N$-th. $L_f$ ($L\leq{L_f}\leq{N}$) is the length of the
sliding window that must be equal or larger than $\emph{L}$ for
smaller inter-symbol interference and larger diversity gain, and
smaller than or equal to $N$. When $L_f=N$, the matched filter
becomes $\textbf{H}^H$. For simplicity, we may rewrite
(\ref{Eq:Hrec}) as
\begin{equation}
    \textbf{y}_{k}=
    \textbf{J}
    \textbf{s}
    +\widetilde{\textbf{n}}_{k},
\label{Eq:PartY}
\end{equation}
where
$\textbf{y}_{k}=\textbf{H}_{k}^H\textbf{r}\in\mathbb{C}^{L_f\times{1}}$,
$\textbf{J}=\textbf{H}_{k}^H\textbf{H}\in\mathbb{C}^{L_f\times{N}}$,
and
$\widetilde{\textbf{n}}_{k}=\textbf{H}_{k}^H\textbf{n}\in\mathbb{C}^{L_f\times{1}}$.
We call this process \emph{horizonal slicing}, since it takes $L_f$
rows of $\textbf{H}$. $\textbf{J}$ is given by
\begin{equation}
        \textbf{J}= \left[\begin{array}{ccc}
     \textbf{h}_k^H\textbf{h}_1&\cdots&\textbf{h}_k^H\textbf{h}_N\\
     \vdots&\ddots&\vdots\\
     \textbf{h}_{\emph{k}+\emph{L}_\emph{f}-1}^\emph{H}\textbf{h}_1&\cdots&\textbf{h}_{\emph{k}+\emph{L}_\emph{f}-1}^\emph{H}\textbf{h}_N\\
      \end{array}\right],
\label{Eq:PartJ}
\end{equation}
where $\textbf{h}_i\in\mathbb{C}^{(N+L-1)\times{1}}$ denotes the
$i$-th column of matrix $\textbf{H}$. The length of the forward
filter has been defined as $L_f$ in (\ref{Eq:Hrec}).

\subsubsection{Decision Feedback Process}
The function of this step is to suppress the effects of the detected
terms.

In order to further decrease the complexity of (\ref{Eq:PartY}), we
can just consider a certain number of the transmitted symbols, and
have
\begin{equation}
    \textbf{y}_{k}
    \approx
    \textbf{J}_{k}
    \textbf{s}_{k}
    +\widetilde{\textbf{n}}_{k},
\label{Eq:APartY}
\end{equation}
where
$\textbf{J}_{k}\in\mathbb{C}^{L_f\times(\emph{k}+\emph{L}_\emph{f}-1)}$
can be constructed by taking the first column to the
$\emph{k}+\emph{L}_\emph{f}-1$-th column of $\textbf{J}$ in
(\ref{Eq:PartJ}), and
$\textbf{s}_{k}=[\emph{s}_1,\ldots,\emph{s}_{\emph{k}+\emph{L}_\emph{f}-1}]^\emph{T}$.
We call this process as \emph{vertical slicing}, since it takes
$k+L_f-1$ columns of $\textbf{J}$. Moreover, (\ref{Eq:APartY}) can
be decomposed with respect to each transmitted symbol
\begin{equation}
    \textbf{y}_{k}
    \approx
    \sum_{\emph{i}=1}^{\emph{k}+\emph{L}_\emph{f}-1}\textbf{j}_\emph{i}\emph{s}_\emph{i}
    +\widetilde{\textbf{n}}_{k},
  \label{Eq:RecForward}
\end{equation}
where $\textbf{j}_\emph{i}\in\mathbb{C}^{\emph{L}_\emph{f}\times1}$
stands for the $\emph{i}$-th column of the matrix $\textbf{J}_{k}$,
and $\emph{s}_\emph{i}$ represents the $\emph{i}$-th transmitted
symbol. Note that (\ref{Eq:RecForward}) is equivalent to
(\ref{Eq:PartY})  when the \emph{vertical slicing} includes all the
symbols in $\textbf{J}$, $L_f=N+1-k$, which implies the length of
$L_f$ will have some effect on the system performance, and the
effect of $\emph{L}_\emph{f}$ will be discussed further in the
performance analysis and simulation results sections. We can write
(\ref{Eq:RecForward}) as
\begin{equation}
    \textbf{y}_{k}
    \approx
    \sum_{\emph{i}=1}^{\emph{k}-1}\textbf{j}_\emph{i}\emph{s}_\emph{i}
    +\textbf{j}_\emph{k}\emph{s}_\emph{k}+\sum_{\emph{i}=\emph{k}+1}^{\emph{k}+\emph{L}_\emph{f}-1}\textbf{j}_\emph{i}s_\emph{i}
    +\widetilde{\textbf{n}}_{k},
  \label{Eq:RecForwardfurther}
\end{equation}
where
$\sum_{\emph{i}=1}^{\emph{k}-1}\textbf{j}_\emph{i}\emph{s}_\emph{i}$
stands for the detected terms that can be rebuilt by the past
decisions, $\textbf{j}_\emph{k}\emph{s}_\emph{k}$ is the current
target, and
$\sum_{\emph{i}=\emph{k}+1}^{\emph{k}+\emph{L}_\emph{f}-1}\textbf{j}_\emph{i}s_\emph{i}$
represents the undetected terms. The function of the \emph{feedback
process} is to reconstruct
$\sum_{\emph{i}=1}^{\emph{k}-1}\textbf{j}_\emph{i}\emph{s}_\emph{i}$
for later interference cancellation. Therefore, it is important to
decide the length of the backward filter, $L_b$. Based on the
expressions of $\textbf{H}$ and (\ref{Eq:PartJ}), we have
$\textbf{j}_1=\textbf{j}_2=\cdots=\textbf{j}_{k-L-2}=\textbf{0}$,
and thus, the \emph{length} of the \emph{backward filter} $L_b$ can
be fixed at $L-1, (\emph{L}>1)$ to reconstruct the effects of past
decisions. (\ref{Eq:RecForwardfurther}) can be rewritten by
simplifying the detected terms
\begin{equation}
    \textbf{y}_{k}
    \approx
    \sum_{\emph{i}={\emph{k}-\emph{L}_\emph{b}}}^{\emph{k}-1}\textbf{j}_\emph{i}\emph{s}_\emph{i}
    +\textbf{j}_\emph{k}\emph{s}_\emph{k}+\sum_{\emph{i}=\emph{k}+1}^{\emph{k}+\emph{L}_\emph{f}-1}\textbf{j}_\emph{i}\emph{s}_\emph{i}
    +\widetilde{\textbf{n}}_{k},
  \label{Eq:RecForwardbackward}
\end{equation}
where $\emph{L}_\emph{b}$ equals $\emph{L}-1$. As in
(\ref{Eq:RecForwardbackward}),
$\sum_{\emph{i}={\emph{k}-\emph{L}_\emph{b}}}^{\emph{k}-1}\textbf{j}_\emph{i}\emph{s}_\emph{i}$
can be reconstructed from past decisions, the following past
decision cancellation process can be applied
\begin{equation}
    \widetilde{\textbf{y}}_{k}=\textbf{y}_{k}
    -\sum_{\emph{i}={\emph{k}-\emph{L}_\emph{b}}}^{\emph{k}-1}\textbf{j}_\emph{i}\emph{s}_\emph{i}.
  \label{Eq:BackwardRestruction}
\end{equation}
The above process is very similar to the decision feedback
cancellation process, but unlike MMSE-DFE, we do not need to
calculate the coefficients of the feedback filter, moreover, the
length of $\emph{L}_\emph{b}$ is fixed at $\emph{L}-1$, which means
that only $\emph{L}-1$ past decisions need to be fed back, which is
much less than what is typically required by MMSE-DFE.
\subsubsection{Approximate ML}
This step aims to achieve near optimal detection by applying the
Gaussian approximation. We substitute (\ref{Eq:RecForwardbackward})
into (\ref{Eq:BackwardRestruction}) and get
\begin{equation}
    \widetilde{\textbf{y}}_{k}= \textbf{j}_\emph{k}s_\emph{k}+
    \sum_{\emph{i}=\emph{k}+1}^{\emph{k}+\emph{L}_\emph{f}-1}\textbf{j}_\emph{i}\emph{s}_\emph{i}
    +\widetilde{\textbf{n}}_{k}.
  \label{Eq:AfterBackwardRestruction}
\end{equation}
In order to decode $\emph{s}_\emph{k}$ with low computational
complexity while maintaining the performance comparable to the ML
decoder, we treat the undetected terms
$\sum_{\emph{i}=\emph{k}+1}^{\emph{k}+\emph{L}_\emph{f}-1}\textbf{j}_\emph{i}\emph{s}_\emph{i}$
and the noise vector
$\widetilde{\textbf{n}}_{\emph{k},\emph{k}+\emph{L}_\emph{f}-1}$
together as a new complex-valued Gaussian vector with matching mean
and covariance matrix, such that (\ref{Eq:AfterBackwardRestruction})
can be expressed as
\begin{equation}
    \widetilde{\textbf{y}}_{k}
    =
    \textbf{j}_\emph{k}s_\emph{k}+
    \textbf{$\boldsymbol{\eta}$}_\emph{k},
  \label{Eq:A_ML}
\end{equation}
where $\boldsymbol{\eta}_\emph{k}$ represents a vector with size
$L_f\times{1}$ of zero-mean complex-valued Gaussian random variables
with covariance
\begin{equation}
\textbf{$\boldsymbol{\Lambda}$}_\emph{k}=
\textbf{J}_{k+1}\textbf{J}_{k+1}^\emph{H}
+\sigma^2\textbf{J}_{k}^{'},
  \label{Eq:GaussAprox}
\end{equation}
where $\textbf{J}_{k+1}$ can be constructed by using the $(k+1)$-th
column as the $(\emph{k}+\emph{L}_\emph{f}-1)$-th column of
$\textbf{J}_k$ and $\textbf{J}_{k}^{'}$ can be obtained by taking
the $k$-th column to the $(\emph{k}+\emph{L}_\emph{f}-1)$-th column
of $\textbf{J}_k$. According to the \emph{central limit theorem},
the \emph{accuracy} of the Gaussian assumption can be improved by
increasing the length of the \emph{forward filter} (sliding window),
$L_f$.

As $\boldsymbol{\eta}_\emph{k}$ has an approximate Gaussian
distribution, the likelihood function
$\emph{p}(\widetilde{\textbf{y}}_{k}|\emph{s}_\emph{k})$ is given by
\begin{align}
    {p(\widetilde{\textbf{y}}_{k}|\emph{s}_\emph{k})}
    \propto
    \texttt{exp}
        \left(
            -(\widetilde{\textbf{y}}_{k}-\textbf{j}_\emph{k}\emph{s}_\emph{k})
    ^\emph{H}\textbf{$\boldsymbol{\Lambda}$}_\emph{k}^{-1}
        \right.&
    \left.(\widetilde{\textbf{y}}_{k}-\textbf{j}_\emph{k}\emph{s}_\emph{k})\right).
\end{align}

Finally, $s_k$ can be recovered by the following \emph{ML} detector
\begin{equation}
    \emph{s}_\emph{k}=
    \underset{\emph{s}_\emph{k}\in\mathcal{A}}{\texttt{arg\hspace{.4em}min}}\left((\widetilde{\textbf{y}}_{k}-\textbf{j}_\emph{k}\emph{s}_\emph{k})
    ^\emph{H}\textbf{$\boldsymbol{\Lambda}$}_\emph{k}^{-1}(\widetilde{\textbf{y}}_{k}-\textbf{j}_\emph{k}\emph{s}_\emph{k})\right),
      \label{Eq:A_ML_decoder}
\end{equation}
at $\emph{k}=\emph{N}-\emph{L}_\emph{f}+1$, there are no more new
received signals outside the sliding window. So, we can then simply
decode each undetected symbol by treating the rest as Gaussian term
and removing the effects of the detected symbols. This decoding
process is very similar to the case of
$\emph{k}<\emph{N}-\emph{L}_\emph{f}+1$ by fixing the sliding
window. The overall A-ML-DFBE algorithm is summarized in Table
\ref{Tab:Algorithm}.

\subsection{Computational Complexity Analysis}
Before we show the complexity comparisons, we present how to further
reduce the proposed equalizer complexity. Note that, in
(\ref{Eq:RecForwardbackward}), the detected terms
$\sum_{\emph{i}={\emph{k}-\emph{L}_\emph{b}}}^{\emph{k}-1}\textbf{j}_\emph{i}\emph{s}_\emph{i}$
can be rewritten as $ \textbf{J}_{k-1}\textbf{s}_{k-L_b}$ with size
$L_f\times1$, where
$\textbf{s}_{k-1}=[s_{\emph{k}-\emph{L}_\emph{b}},\ldots,{s_{\emph{k}-1}}]^\emph{T}$
has size $\emph{L}_\emph{b}\times1$. With respect to the diagonal
element $\textbf{h}_g^H\textbf{h}_g$ in $\textbf{J}$, when $g>L$, we
can find that
\begin{align}
    \textbf{h}_g^H\textbf{h}_i=0, i\geq{g+L},
    \nonumber \\
    \textbf{h}_i^H\textbf{h}_g=0, i\leq{g-L},
    \label{Eq:eleinJ}
\end{align}
and thus, $\textbf{J}_{k-1}$ has the following form
$$
        \textbf{J}_{k-1}= \left[\begin{array}{cccc}
     \textbf{h}_\emph{k}^\emph{H}\textbf{h}_{\emph{k}-\emph{L}_\emph{b}}&\cdots&\cdots&\textbf{h}_\emph{k}^\emph{H}\textbf{h}_{\emph{k}-1}\\
     0&\textbf{h}_{\emph{k}+1}^\emph{H}\textbf{h}_{\emph{k}-\emph{L}_\emph{b}+1}&\ldots&\vdots\\
     \vdots&\ddots&\ddots&\vdots\\
     0&\cdots&0&\textbf{h}_{k+L_b-1}^H\textbf{h}_{k-1}\\
     0&\cdots&\cdots&0\\
     \vdots&\ddots&\ddots&\vdots\\
     0&\cdots&\cdots&0\\
      \end{array}\right],
$$
which has size $L_f\times{L_b}$. We can observe that there are only
$\sum_{i=1}^{L_b}i=\frac{1+L_b}{2}L_b$ non-zero elements in
$\textbf{J}_{k-1}$ so that the reconstruction of the detected terms
$\sum_{i=k-L_b}^{k-1}\textbf{j}_is_i$ can be further simplified.
Similarly, in (\ref{Eq:A_ML_decoder}), the calculation of
$\boldsymbol{\Lambda}_k$ and $\textbf{j}_k$ can be simplified as
well.

Now, we discuss the complexity of the A-ML-DFBE, linear-MMSE
\cite{Simon-Adaptive-Filter}, MMSE-DFE \cite{Simon-Adaptive-Filter},
and BAD \cite{IEEE-BAD} detectors in terms of the number of
additions and multiplications. The resulting values are given in
Table \ref{Tab:Complexity}, obtained by inspection of the relevant
algorithms in
Table~\ref{Tab:Algorithm},~\cite{Simon-Adaptive-Filter},
and~\cite{IEEE-BAD}. Details of the computation of complexity, for
example the matrix inversion, can be found
in\cite{Golub-Matrix-Computations}. The computational complexity of
the A-ML-DFBE algorithm is a function of the frame length~($N$), the
impulse response length~($L$), and the length of the forward
filter~($L_f$), which is obtained on the basis of
Table.~\ref{Tab:Algorithm}. From the table, we observe that
A-ML-DFBE has the same order of complexity as the linear-MMSE and
MMSE-DFE. But A-ML-DFBE is less complex than MMSE-DFE since the
A-ML-DFBE requires smaller $L_f$ value, and it does not require to
build up the backward filter. In comparison to linear-MMSE, the
A-ML-DFBE needs relatively even shorter forward filter and thus has
lower complexity. The relation between the filter length and the
performance can be clearly observed in the simulation results
section. BAD requires complexity at least double of MMSE-DFE. Note
that with regard to computational complexity, we focus on
time-domain implementation even though a low-complexity
frequency-domain implementation is also possible by making use of
the block-circulant structure that can be created by the guard
interval. In addition, note that the matrix inversion lemma can be
used to reduce the complexity from cubic to quadratic order, but it
does not affect the above conclusions.
\section{Performance Analysis}
\subsection{Analytical SER and BER Derivations}
In this subsection, we analyze the symbol error rate (SER) as well
as the bit error rate (BER) performance of the A-ML-DFBE. Note that
the tail detection only contains the operation of very few symbols,
and thus, the performance is dominated by Step 2 of the A-ML-DFBE
process in Table \ref{Tab:Algorithm}, which will now be analyzed. We
assume that all the decisions are accurate for analysis, which is a
normal assumption in decision feedback theory
\cite{Simon-Adaptive-Filter}. In (\ref{Eq:A_ML}), which contains
correlated noise, $\textbf{$\boldsymbol{\eta}$}_\emph{k}$, the
pre-whitening filter,
$\boldsymbol{\Psi}_k=\boldsymbol{\Lambda}_k^{-\frac{1}{2}}$, can be
applied to make the variance of the noise uncorrelated
\begin{equation}
    \boldsymbol{\Psi}_k\widetilde{\textbf{y}}_{k}=\boldsymbol{\Psi}_k\textbf{j}_\emph{k}s_\emph{k}+
    \boldsymbol{\Psi}_k\textbf{$\boldsymbol{\eta}$}_\emph{k},
\end{equation}
where $\boldsymbol{\Psi}_k\textbf{$\boldsymbol{\eta}$}_\emph{k}$
with size $L_f\times{1}$ has a Gaussian distribution with zero mean
and all components have unit variance.

Since the noise now has become white Gaussian, the matched filter,
$(\boldsymbol{\Psi}_k\textbf{j}_k)^H$, can be employed and we have
the following received signal equation in scalar form
\begin{equation}
    y_{k}^{'}
    =
    \xi_k{s_k}
    +
    \upsilon_k,
 \label{Eq:MP}
\end{equation}
where
$y_{k}^{'}=\left(\boldsymbol{\Psi}_k\textbf{j}_k\right)^H\boldsymbol{\Psi}_k\widetilde{\textbf{y}}_{k}$,
$\xi_k=\|\boldsymbol{\Psi}_k\textbf{j}_k\|^2$, and
$\upsilon_k=\left(\boldsymbol{\Psi}_k\textbf{j}_k\right)^H\boldsymbol{\Psi}_k{\boldsymbol{\eta}}_k$,
which is a scalar with zero mean and variance
$\|\boldsymbol{\Psi}_k\textbf{j}_k\|^2$. The SER for $M$-PSK
constellation is given by \cite{Simon-Digital-Comms}
\begin{equation}
    {\texttt{SER}}_M^k=\frac{1}{\pi}\int_0^{\frac{(M-1)\pi}{M}}
    \texttt{exp}\left(-\frac{g_\texttt{psk}\gamma_k}{\sin^2\theta}\right)d\theta,
 \label{Eq:SERBPSKkth}
\end{equation}
where $g_\texttt{psk}\triangleq\sin^2\frac{\pi}{M}$,
$\gamma_k\triangleq\frac{|\xi_k{s_k}|^2}{\texttt{var}(\upsilon_k)}=\frac{\xi_k^2}{(\boldsymbol{\Psi}_k\textbf{j}_k)^H(\boldsymbol{\Psi}_k\textbf{j}_k)}=\|\boldsymbol{\Psi}_k\textbf{j}_k\|^2$,
and $M$ denotes the constellation size. The average BER for
$M$-\texttt{PSK} can be written as:
\begin{equation}
    {{\texttt{BER}}_M}=\frac{1}{N-L_f+2}\sum_{k=1}^{N-L_f+2}{\texttt{BER}}_M^k,
 \label{Eq:SERBPSKVBLAST}
\end{equation}
where
${\texttt{BER}}_M^k\approx\frac{1}{\texttt{log}_2M}{\texttt{SER}}_M^k$
\cite{Proakis-Digital-Comms} for high SNR and Gray mapping. Since
the tail is normally short, which has length $L_f-2$, in comparison
to the whole frame length $N$, hence its effects can be neglected.
Note that in time-invariant channel,
$\texttt{SER}_M^1=\texttt{SER}_M^2=\cdots=\texttt{SER}_M^{N-L_f+2}$
due to the property of $\textbf{J}$
($\gamma_1=\gamma_2=\cdots=\gamma_{N-L_f+2}$) by assuming perfect
decision feedback at high SNR.

\subsection{Multipath Diversity Analysis}
Next, we analyze further the behavior of the proposed A-ML-DFBE at
high SNR. Assuming perfect channel estimation at the receiver, and
taking (\ref{Eq:SERBPSKkth}) as an example, it can be upper bounded
by~\cite{Proakis-Digital-Comms}
\begin{align}
    \texttt{SER}_M^k
    \leq
    \frac{1}{2}\texttt{exp}\left(-\frac{g_\texttt{psk}}{\sin^2\theta}{\gamma_k}\right)
    &\approx
    \frac{1}{2}\texttt{exp}\left(-\frac{g_\texttt{psk}}{\sigma^{2}L_f\sin^2\theta}{\sum_{i=0}^{\texttt{min}(L_f-1,L-1)}|h_i(t)|^2}\right),
    \label{Eq:upperbound}
\end{align}
where ${\gamma_k}\approx
    \frac{1}{\sigma^{2}L_f}\sum_{i=0}^{\texttt{min}(L_f-1,L-1)}|h_i(t)|^2$ at high SNR (Refer to Appendix I for the derivation). In order to obtain good performance in terms of multipath
combining and inter-symbol interference suppression, we should
choose $L_f\geq{L}$.
Then, by averaging (\ref{Eq:upperbound}) over the Rayleigh PDF
\cite{Tarokh-STTC}, equation (\ref{Eq:upperbound}) becomes
\begin{equation}
    \overline{\texttt{SER}_M^k}
    \leqslant
    \frac{1}{2}\left(\frac{g_\texttt{psk}\texttt{SNR}}{L{\cdot}L_f\sin^2\theta}\right)^{-L},
      \label{Eq:finalbound}
\end{equation}
which indicates that the A-ML-DFBE achieve the maximum multipath
diversity order $L$.
\subsection{Analysis of the Length of the Forward Filter (Sliding
Window) and Backward Filter}
It has been shown that the forward filter length, $L_f$, is a very
important parameter in the proposed A-ML-DFBE. In this subsection,
we discuss the behaviors of $L_f$: 1) Increasing the value of $L_f$
can improve the robustness of (\ref{Eq:A_ML_decoder}) due to the
following reasons: Firstly, as shown in (\ref{Eq:PartJ}), larger
value of $L_f$ can incorporate more received signals as well as
channel information in the forward filter; Secondly, indicated by
(\ref{Eq:GaussAprox}), increasing $L_f$ can make the Gaussian
assumption more accurate; 2) While the performance can be enhanced,
as shown in Table \ref{Tab:Complexity}, the complexity will
correspondingly go up. Hence, for A-ML-DFBE, a complexity and
performance tradeoff can be realized by adjusting $L_f$; 3)
Performance gets converged by increasing the value of $L_f$ as the
Gaussian assumption becomes accurate enough. This implies that
moderate length of the forward filter can deliver good performance;
4) Given by Subsection-IV-B, $L_f$ should be equal or larger than
$L$ for maximum diversity order; 5) The length of the backward,
$L_b$, always equals $L-1$ due to the special structure of
$\textbf{H}$.
\subsection{Analysis of the Matched Filter in (\ref{Eq:Hrec})}
Note that the matched filter in (\ref{Eq:Hrec}) can obtain some
additional information from the received signals outside the slicing
window. Recalling (\ref{Eq:PartY})--(\ref{Eq:RecForwardbackward}),
$\textbf{y}_{k}$ can be written as
\begin{equation}
    \textbf{y}_{k}=[\textbf{h}_k^H\textbf{r},\ldots,\textbf{h}_{k+L_f-1}^H\textbf{r}]^T.
\end{equation}
Although some information is lost after horizonal and vertical
slicing, some gains can be still realized by considering the whole
received signal, $\textbf{r}$.

Supposing the matched filter is removed, the detection procedures in
Table \ref{Tab:Algorithm} can be used but it will lead to
performance degradation since only the received signals inside the
sliding window will be considered, where
$\textbf{y}_{k}=[r_k,\ldots,r_{k+L_f-1}]^T$.
As a result, the length of the forward filter has to be increased to
make up the performance loss caused by the slicing processes in
order to obtain the same performance. Note also if the length of the
forward filter is equal to $N$, the A-ML-DFBE directly enters the
'Tail Detection' step (Step \textbf{3}) in Table
\ref{Tab:Algorithm}, which will make no difference in performance
whether or not the matched filter is used since there is no slicing
operations at all. However, the value of $L_f$ is normally much less
than $N$. Theoretically, using the same methods as shown in Appendix
I, it is easy to obtain the SNR for the A-ML-DFBE when the matched
filter is removed. Due to the space limitation, we drop the detailed
derivation part. But we can conclude that the performance of
A-ML-DFBE can be upper-bounded by the same equalize without using
the matched filter.
\section{Simulation Results}
In all simulations, BPSK constellation is used to generate a rate
1bps/Hz transmission. We plot the BER versus the signal-to-noise
ratio~(SNR). For analytical results, we assume perfect decision
feedback, but for simulated results we use the feedback decisions.
The performance is determined over doubly selective Rayleigh fading
channels. The impulse response length is $L=5$, and, thus, the
length of the backward filter of the A-ML-DFBE can be fixed as
$L_b=L-1=4$. Jakes' Model is applied to construct time-selective
Rayleigh fading channel for each subpath. The carrier frequency
$f_c=2$ GHz and the symbol period $T_s=128/c$, where $c$ is the
speed of light. The simulation results are plotted with two speeds:
$v=5$ $vkm/h$ and $150$ $km/h$ (corresponding to $f_dT_s=0.0001$ and
$0.0093$, where Doppler frequency $f_d=vf_c/c$). The frame length
$N$ is 128.

\begin{figure}[]
\centering
\includegraphics[height=4.0in,width=6.0in]{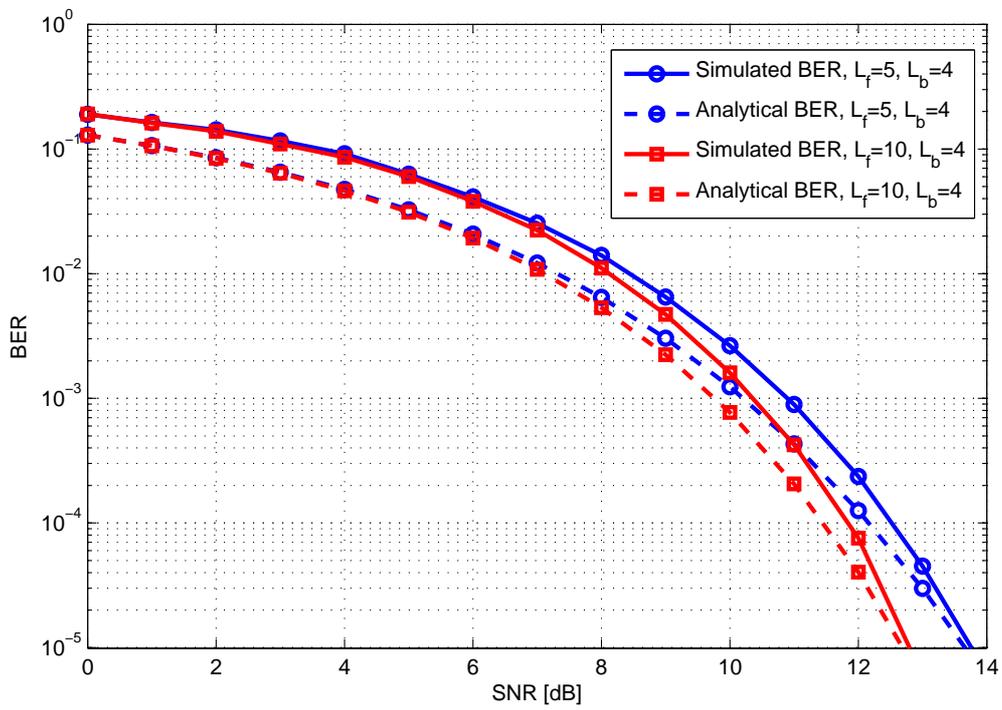}
\caption{Analytical BER performance of the A-ML-DFBE over a doubly
selective fading channel with perfect channel estimation ($L=5$,
$f_dT_s=0.0001$) and simulated BER.} \label{Fig:AnalyticalBER}
\end{figure}

\begin{figure}[]
\centering
\includegraphics[height=4.0in,width=6.0in]{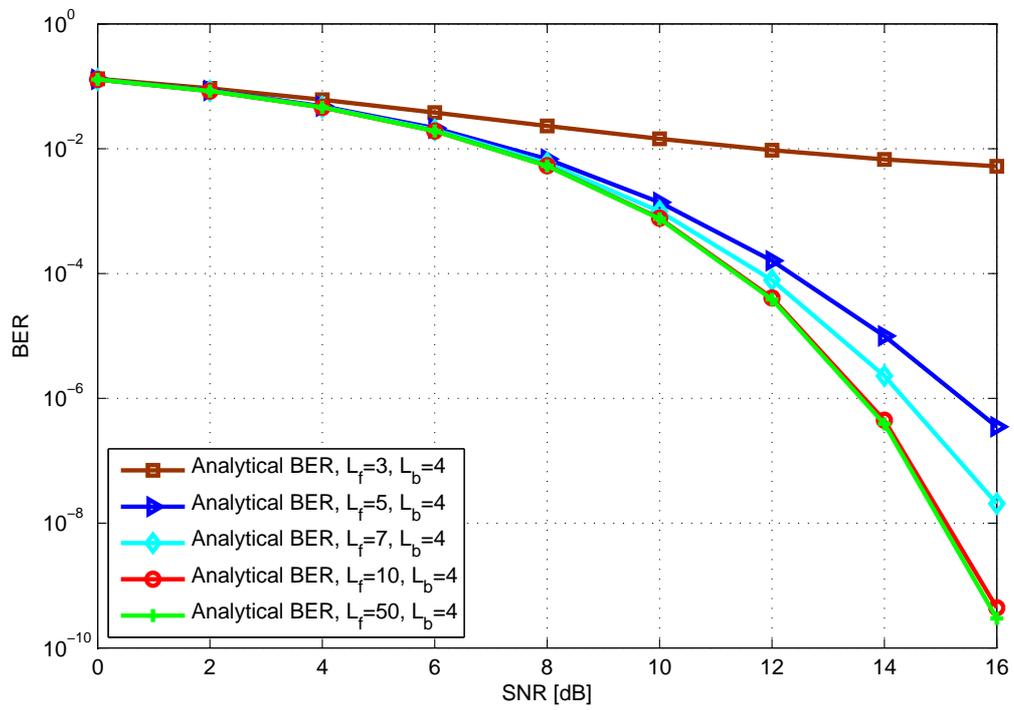}
\caption{Analytical BER performance of the A-ML-DFBE with various
forward filter length over a doubly selective fading channel with
perfect channel estimation ($L=5$, $f_dT_s=0.0001$).}
\label{Fig:AnalyticalBERdifftaps}
\end{figure}

In Fig. \ref{Fig:AnalyticalBER} and Fig.
\ref{Fig:AnalyticalBERdifftaps}, we examine the analytical BER
performance obtained in (\ref{Eq:SERBPSKVBLAST}) assuming that the
channel estimation is perfect. The simulations are plotted with the
vehicle speed: $v=5 km/h$. In Fig. \ref{Fig:AnalyticalBER}, we
compare the analytical BER with the simulated BER. It can be
observed that the analytical BER is close and asymptotically
converges to the simulated curves at high SNR. In Fig.
\ref{Fig:AnalyticalBERdifftaps}, the analytical BER for A-ML-DFBE is
plotted employing different forward filter lengths. As discussed
earlier, the length of the forward filter, $L_f$, should be at least
equal to $L$ in order to realize good performance. From Fig.
\ref{Fig:AnalyticalBERdifftaps}, we can see that the proposed
A-ML-DFBE with $L_f=5$ provides much better performance than that
with $L_f=3$, and as the value of $L_f$ increases, the performance
begins to converge. It can be also seen that for A-ML-DFBE, $L_f=10$
(two times $L$) is enough to obtain good BER performance.

\begin{figure}[]
\centering
\includegraphics[height=4.0in,width=6.0in]{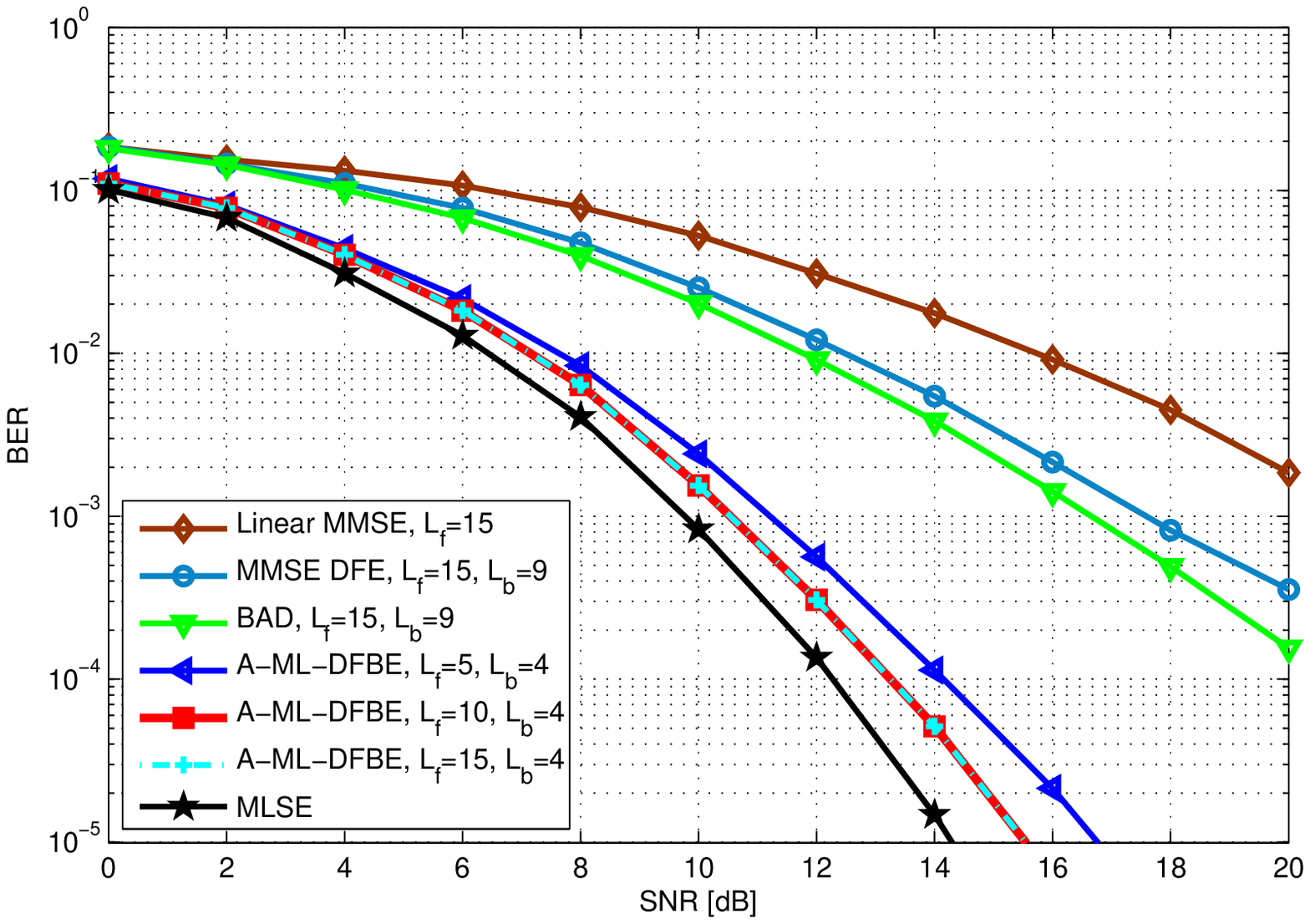}
\caption{Simulated BER performance of the A-ML-DFBE over a doubly
selective fading channel with LS channel estimation ($L=5$,
$f_dT_s=0.0001$). Shown for comparisons are the Linear MMSE,
MMSE-DFE, BAD, and MLSE.} \label{Fig:RandomH}
\end{figure}

In Fig. \ref{Fig:RandomH}, simulation results for the A-ML-DFBE
detector are illustrated in comparison with conventional linear
MMSE, MMSE-DFE, BAD, and MLSE decoders. The simulations are plotted
with the vehicle speed: $v=5$ $km/h$. Least square~(LS) channel
estimation~\cite{Simon-Adaptive-Filter} is used. From Fig.
\ref{Fig:RandomH}, it can be observed that at BER=$10^{-3}$, the
performance of A-ML-DFBE with $\emph{L}_\emph{f}=5$ is far better
than the linear MMSE and the MMSE-DFE equalizers. There is only 2 dB
loss compared to the MLSE decoder at BER=$10^{-5}$. At
$\emph{L}_\emph{f}=10$, there is about 0.8 dB loss compared to MLSE.
Almost no difference can be observed for A-ML-DFBE when
$\emph{L}_\emph{f}$ is increased to 15 since $L_f=10$ is sufficient
to make the performance converge. Note that when
$\emph{L}_\emph{f}=15$, A-ML-DFBE gives almost the same performance
as $\emph{L}_\emph{f}=10$, which demonstrates that only a small
value of $\emph{L}_\emph{f}$ is required to achieve good
performance. We can also see that A-ML-DFBE with $L_f=5$ can provide
much better performance than BAD with $L_f=15$. Note that our
A-ML-DFBE has lower complexity than MMSE-DFE, and thus, lower than
BAD. Clearly, from Fig. \ref{Fig:AnalyticalBER} to Fig.
\ref{Fig:RandomH},we can see that there exists a complexity and
performance tradeoff in terms of $L_f$. Performance can be improved
by increasing the length of the forward filter (slicing window). In
addition, performance convergence can be also observed, which
indicates that limited value of $L_f$ is enough to deliver most of
the performance gain.

\begin{figure}[]
\centering
\includegraphics[height=4.0in,width=6.0in]{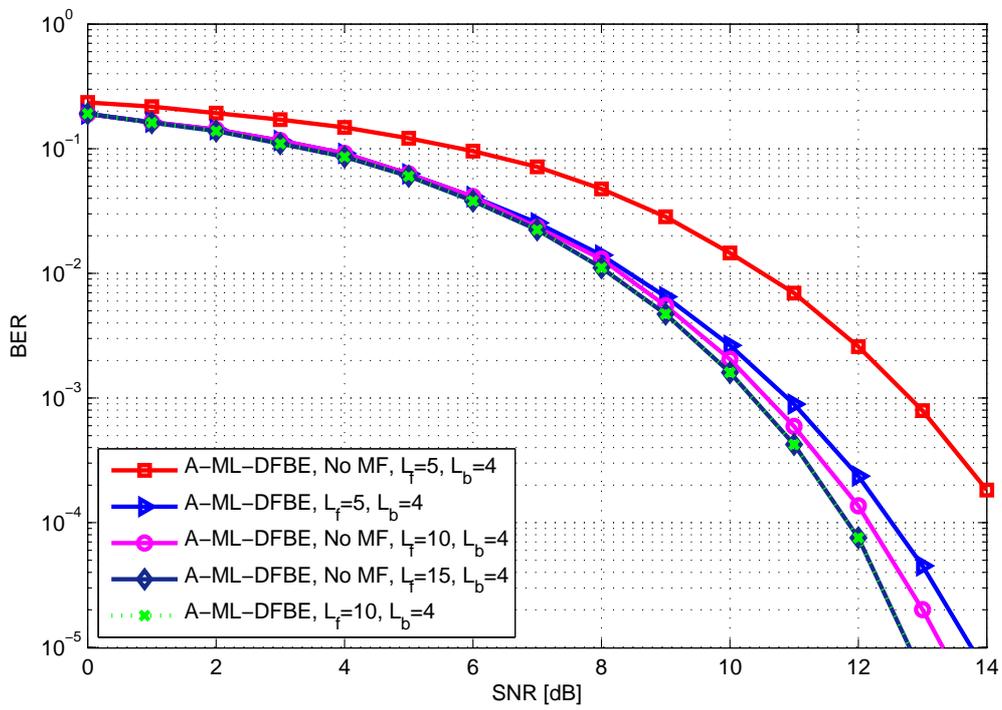}
\caption{Simulated BER performance of the A-ML-DFBE over a doubly
selective fading channel with perfect channel estimation ($L=5$,
$f_dT_s=0.0001$). Shown for comparisons are A-ML-DFBE with and
without matched filter (MF).} \label{Fig:Matchedfilter}
\end{figure}

In Fig. \ref{Fig:Matchedfilter}, simulation comparisons are made for
A-ML-DFBE without using the matched filter. Perfect channel
estimation is assumed. Vehicle speed, $v=5 km/h$, is adopted. We
choose different $L_f$ values for the no matched filter case: 5, 10,
and 15 and $L_b$ remains the same: 4. It is shown that at $L_f=5$,
the performance without the matched filter is worse than with it. We
can also observe the significant performance loss due to the small
value of $L_f$. It is shown that $L_f$ must be 15 for the system
with no matched filter to provide the same performance as the
matched filter system with $L_f=10$. Hence, from the simulation
results we can see that the matched filter is very important for
system performance. Note that as discussed in the complexity
analysis part, Subsection III-B, the forward and backward filter
taps are actually fixed and can be obtained before the A-ML-DFBE
detection. The complexity increase by the use of the matched filter
is much more worthwhile than to increase the length of the forward
filter without using the matched filter.

\begin{figure}[]
\centering
\includegraphics[height=4.0in,width=6.0in]{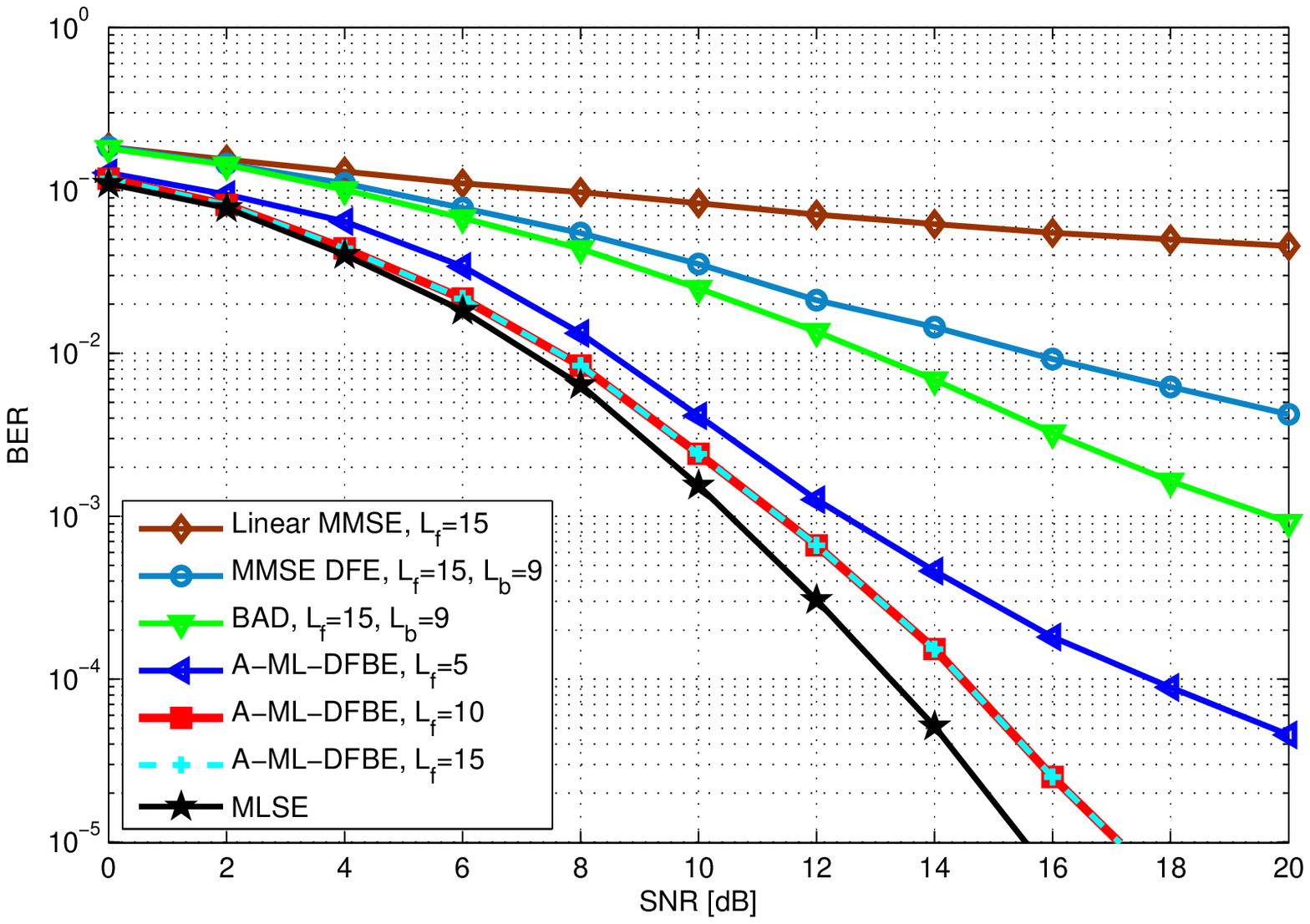}
\caption{Performance of the A-ML-DFBE over a doubly selective fading
channel with LS channel estimation ($L=5$, $f_dT_s=0.0093$). Shown
for comparisons are the Linear MMSE, MMSE-DFE, BAD, and MLSE.}
\label{Fig:TimeVariantCHE}
\end{figure}

In Fig.~\ref{Fig:TimeVariantCHE}, simulation results for the
A-ML-DFE detector are illustrated in comparison with conventional
linear MMSE, MMSE-DFE, BAD, and MLSE decoders using LS channel
estimation and the vehicle speed is $v=150 km/h$. Here, we choose
different values for $\emph{L}_\emph{f}$ for A-ML-DFBE. From the
simulation results, we can still observe that the performance of
A-ML-DFBE converged at $L_f=10$, and no gain can be obtained at
$L_f=15$. Due to the time-variant effects, the performance is
degraded compared to the results in Fig.~\ref{Fig:RandomH}. We can
see about 1 dB loss between MLSE and A-ML-DFBE with $L_f=10$ when
BER=$10^{-5}$. However, the proposed equalizer can still
substantially outperform Linear MMSE and MMSE-DFE in all SNR regime.
Around 8 dB performance gain can be obtained by the proposed scheme
with $L_f=5$ compared to the BAD at BER=$10^{-3}$.
\section{Conclusions}
\label{sec:Conclusion} In this paper, we have proposed a simple
approximate ML decision feedback equalizer for doubly selective
fading environment. From the analytical and simulation results, we
conclude that the A-ML-DFBE significantly outperforms the linear
MMSE, MMSE-DFE, and BAD detectors, and provides performance very
close to MLSE. We have shown that when $L_f$ is large enough,
further increases in $L_f$ do not improve performance much. This
implies that the proposed equalizer is quite robust against ISI. A
tradeoff in terms of the complexity and the performance can be
achieved by adjusting the value of $L_f$. Computational complexity
comparison has demonstrated that the A-ML-DFBE requires fewer
additions and multiplications than MMSE based schemes. In addition,
the implementation of the matched filter is very important and the
A-ML-DFBE obtains maximum diversity order when $L_f\geq{L}$.

Due to the DFE processing, parallel computing is difficult to
achieve for the proposed equalizer. However, by adjusting the size
of the data block or the filters (back and forward), or both, the
latency can be reduced. The proposed equalizer can be easily used
for radar communication systems as it is suitable to solve
time-domain equalization problems. In current wireless systems like
UMTS, HSDPA or HSUPA, the A-ML-DFBE can be used to recover signals
similar to MMSE or MMSE-DFE. For LTE or LTE advance, the proposed
algorithm can be extended to realize frequency-domain equalizations.

\appendices
\section{Derivation of Closed-Form Expression of ${\gamma_k}$ at High SNR}
Now, the closed-form expression of ${\gamma_k}$ at high SNR is
derived in terms of $L_f$ and $L$. From Subsection IV-A,
${\gamma_k}$ can be written as
\begin{align}
    {\gamma_k}
    =
    \|\boldsymbol{\Psi}_k\textbf{j}_k\|^2
    &=
    \textbf{j}_k^H\boldsymbol{\Lambda}_{k}^{-1}\textbf{j}_k
    =
    \textbf{j}_k^H\left(\sigma^2\textbf{X}+\textbf{Y}\textbf{Y}^H
    \right)^{-1}\textbf{j}_k,
    \label{EQ:distance1kk1}
\end{align}
where for convenience $\textbf{Y}\triangleq\textbf{J}_{k+1}$ has
size $L_f\times{L_f-1}$, and
$\textbf{X}\triangleq\textbf{J}_{k}^{'}$. By using the Kailath
Variant
$(\textbf{A}+\textbf{BC})^{-1}=\textbf{A}^{-1}-\textbf{A}^{-1}\textbf{B}(\textbf{I}+\textbf{CA}^{-1}\textbf{B})^{-1}\textbf{CA}^{-1}$
\cite{Horn-Matrix-Analysis}, the inversion term on the right side of
(\ref{EQ:distance1kk1}) can be further written as
\begin{align}
    {\gamma_k}
    =
    \sigma^{-2}\textbf{X}^{-1}
    -
    \sigma^{-2}\textbf{X}^{-1}\textbf{Y}\left(\sigma^{2}\textbf{I}_{L_f-1}+\textbf{Y}^H\textbf{X}^{-1}\textbf{Y}\right)^{-1}
    \textbf{Y}^H\textbf{X}^{-1}.
    \label{EQ:distance1kk}
\end{align}

At high SNR, as $\sigma^{2}\rightarrow{0}^+$, the effect of
$\sigma^{2}\textbf{I}_{L_f-1}$ is comparatively small, which can be
ignored from an asymptotic point of view. Hence, we have the
following approximation for the second term in
(\ref{EQ:distance1kk})
\begin{align}
    \sigma^{-2}\textbf{X}^{-1}\textbf{Y}(\textbf{Y}^H\textbf{X}^{-1}&\textbf{Y})^{-1}
    \textbf{Y}^H\textbf{X}^{-1}
    =\sigma^{-2}
    \textbf{X}^{-\frac{1}{2}}
    \textbf{Z}\left(\textbf{Z}^H\textbf{Z}\right)^{-1}
    \textbf{Z}^H
    \textbf{X}^{-\frac{1}{2}}.
    \label{EQ:distance1kk2}
\end{align}
where $\textbf{Z}\triangleq\textbf{X}^{-\frac{1}{2}}\textbf{Y}$ with
size $L_f\times{(L_f-1)}$, and $(\cdot)^{-\frac{1}{2}}$ represents
the unique positive definite Hermitian root
\cite{Horn-Matrix-Analysis}.

Let $\textbf{Z}^+$ be the Moore-Penrose inverse of matrix
$\textbf{Z}$, and
$\textbf{Z}^+=(\textbf{Z}^H\textbf{Z})^{-1}\textbf{Z}^H$ of size
$(L_f-1)\times{L_f}$. Note that
$\texttt{rank}(\textbf{Z}\textbf{Z}^+)=\texttt{rank}(\textbf{Y})=L_f-1$
and $\textbf{Z}\textbf{Z}^+$ has size $L_f\times{L_f}$. By
eigenvalue decomposition, we can get
$\textbf{Z}\textbf{Z}^+=\textbf{U}\boldsymbol{\Pi}\textbf{U}^H$
where $\textbf{U}$ is the unitary eigenvector matrix and
$\boldsymbol{\Pi}\triangleq\texttt{diag}\{\lambda_1,\ldots,\lambda_{L_f-1},0\}$.
From the definition of $\textbf{Z}\textbf{Z}^+$, we have
$(\textbf{Z}\textbf{Z}^+)^2=\textbf{Z}\textbf{Z}^+$. Therefore,
$\textbf{Z}\textbf{Z}^+$ is idempotent~\cite{Horn-Matrix-Analysis},
and any idempotent matrix has eigenvalue 1 or 0, and thus
$\boldsymbol{\Pi}=\texttt{diag}\{1,\ldots,1,0\}$. We can then get
\begin{align}
    \textbf{j}_k^H\textbf{X}^{-\frac{1}{2}}
    \textbf{Z}\left(\textbf{Z}^H\textbf{Z}\right)^{-1}\textbf{Z}^H
    \textbf{X}^{-\frac{1}{2}}\textbf{j}_k
    &=
    \textbf{j}_k^H\textbf{X}^{-\frac{1}{2}}
    \textbf{U}\boldsymbol{\Pi}\textbf{U}^H
    \textbf{X}^{-\frac{1}{2}}\textbf{j}_k
    \approx
    \frac{L_f-1}{L_f}\textbf{j}_k^H\textbf{X}^{-1}\textbf{j}_k.
    \label{EQ:distance1kk3}
\end{align}
From (\ref{EQ:distance1kk1}), (\ref{EQ:distance1kk}),
(\ref{EQ:distance1kk2}), and (\ref{EQ:distance1kk3}), at high SNR,
we can obtain
\begin{align}
    {\gamma_k}
    \approx
    \frac{1}{\sigma^{2}L_f}\textbf{j}_k^H\textbf{X}^{-1}\textbf{j}_k,
    \label{Eq:fin1}
\end{align}
From (\ref{Eq:PartJ}), we can get
\begin{equation}
    \textbf{j}_k^H\textbf{X}^{-1}\textbf{j}_k
    =
    \textbf{h}_k^H
    \textbf{H}_{k}(\textbf{H}_{k}^H\textbf{H}_{k})^{-1}\textbf{H}_{k}^H
    \textbf{h}_k,
    \label{Eq:fin2}
\end{equation}
where $\textbf{j}_k=\textbf{H}_{k}^H\textbf{h}_k$ and
$\textbf{X}=\textbf{H}_{k}^H\textbf{H}_{k}$ has size
$L_f\times{L_f}$ and rank $L_f$. Since
$\textbf{H}_{k}(\textbf{H}_{k}^H\textbf{H}_{k})^{-1}\textbf{H}_{k}^H$
has the same structure as
$\textbf{Z}\left(\textbf{Z}^H\textbf{Z}\right)^{-1}\textbf{Z}^H$ in
(\ref{EQ:distance1kk3}), we can get the corresponding eigenvalues as
\begin{align}
    \texttt{EIG}\left(\textbf{H}_{k}(\textbf{H}_{k}^H\textbf{H}_{k})^{-1}\textbf{H}_{k}^H\right)
    =
    \texttt{diag}\{\underbrace{0,\ldots,0}_{k-1},\underbrace{1,\ldots,1}_{L_f},\underbrace{0,\ldots,0}_{N+L-L_f-k}\}.
    \label{Eq:eig}
\end{align}
Finally, combining (\ref{Eq:fin1}) and (\ref{Eq:fin2}), at high SNR
as $\sigma^{2}\rightarrow{0}^+$, finally we have
\begin{align}
    \gamma_k
    \approx
    \frac{1}{\sigma^{2}L_f}\sum_{i=0}^{\texttt{min}(L_f-1,L-1)}|h_i(t)|^2.
\end{align}


\pagebreak
\begin{table}
  \renewcommand{\baselinestretch}{1}
  {\caption{\label{Tab:Algorithm} Approximate Maximum Likelihood
    Decision Feedback Block Equalization Algorithm.}}
  \vspace*{-1.5em}
  \small
  \begin{algorithm}
      \item[Initialization of A-ML-DFBE]
        \begin{subalgorithm}
            1. Set $L_f$ ($L_f\geq{L}$).
            \\
            2. Fix $L_b=L-1$.
            \\
        \end{subalgorithm}
    \item[A-ML-DFBE Detection]
        \begin{subalgorithm}
            \textbf{For $k=1:N-L_f$}
                \\
                \indentcolumn{1. \texttt{Forward Process}: Apply the temporal sub matched filter according to
                (\ref{Eq:PartY}).}
                \indentcolumn{2. \texttt{Decision Feedback Process}: Remove the effects
                            reconstructed by the past \indentcolumn {decisions using
                            (\ref{Eq:BackwardRestruction}).}}
                \indentcolumn{3. \texttt{Approximate ML}: Recover the transmitted signals with (\ref{Eq:A_ML_decoder}).}
            \textbf{end;}
        \end{subalgorithm}
            \item[Tail Detection]
        \begin{subalgorithm}
            1. Fix the sliding window:
            \indentcolumn{$\textbf{y}_{N-L_f+1}=[y_{N-L_f+1},\ldots,{y_{N}}]^T$.}
            2. Signal Recovery:
        \end{subalgorithm}
        \begin{subalgorithm}
            \textbf{For} $k=N-L_f+1:N$
                \indentcolumn{1. \texttt{Decision Feedback Process}: Remove the effects
                of the past decisions \indentcolumn
                {using~(\ref{Eq:BackwardRestruction}).}}
                \indentcolumn{2. \texttt{Approximate ML}: Recover the transmitted signals with (\ref{Eq:A_ML_decoder}).}
            \textbf{end;}
        \end{subalgorithm}
  \end{algorithm}
  \vspace{-7mm}
\end{table}

\clearpage
\begin{table}[p]
{\caption{\label{Tab:Complexity} Computational complexity of various
schemes for one sliding window with length $\emph{N}$; $\emph{L}$ is
the number of paths; BPSK~constellations; $\emph{L}_\emph{f}$ and
$\emph{L}_\emph{b}$ stand for the length of the forward and backward
filters, respectively.}}
\begin{center}
\begin{tabular}{|c|c|c|}  \hline
{\textbf{Detector}} & {\texttt{Additions}} & {\texttt{Multiplications}}\\
\hline \texttt{A-ML-DFBE}
        & $\emph{N}[8\emph{L}_\emph{f}^3+34\emph{L}_\emph{f}^2+(6\emph{L}+7)\emph{L}_\emph{f}+(3\emph{L}-1)]$
        & $\emph{N}[(2\emph{L}_\emph{f}^3+42\emph{L}_\emph{f}^2)-(12\emph{L}+19)\emph{L}_\emph{f}+18]$  \\
\hline \texttt{Linear-MMSE}
        & $\emph{N}[8\emph{L}_\emph{f}^3+30\emph{L}_\emph{f}^2+2(3\emph{L}+2)\emph{L}_\emph{f}]$
        & $\emph{N}[2\emph{L}_\emph{f}^3+42\emph{L}_\emph{f}^2-(12\emph{L}-17)\emph{L}_\emph{f}-(6\emph{L}-1)]$\\
\hline \texttt{MMSE-DFE}
        & $\emph{N}[8(\emph{L}_\emph{f}^3+\emph{L}_\emph{b}^3)+42(\emph{L}_\emph{f}^2+\emph{L}_\emph{b}^2)+2(3\emph{L}+2)(\emph{L}_\emph{f}+\emph{L}_\emph{b})]$
        & $\emph{N}[2(\emph{L}_\emph{f}^3+\emph{L}_\emph{b}^3)+42(\emph{L}_\emph{f}^2+\emph{L}_\emph{b}^2)+(12\emph{L}-11)(\emph{L}_\emph{f}+\emph{L}_\emph{b})+6]$\\
\hline \texttt{BAD} &
$\emph{N}[16(\emph{L}_\emph{f}^3+\emph{L}_\emph{b}^3)+84(\emph{L}_\emph{f}^2+\emph{L}_\emph{b}^2)+4(3\emph{L}+2)(\emph{L}_\emph{f}+\emph{L}_\emph{b})]$
        & $\emph{N}[4(\emph{L}_\emph{f}^3+\emph{L}_\emph{b}^3)+84(\emph{L}_\emph{f}^2+\emph{L}_\emph{b}^2)+2(12\emph{L}-11)(\emph{L}_\emph{f}+\emph{L}_\emph{b})+12]$\\
\hline
\end{tabular}
\end{center}
\end{table}
\clearpage
\clearpage
\clearpage
\clearpage
\clearpage
\end{document}